\documentclass[final,5p,times,twocolumn,sort&compress]{elsarticle}

\usepackage{epsfig}

\usepackage{amssymb}
\usepackage{amsmath}

\journal{Physics Letters A}

\begin{document}

\begin{frontmatter}



\title{Macroscopic ground-state degeneracy and magnetocaloric effect in the exactly
\linebreak
 solvable
spin-1/2 Ising-Heisenberg double-tetrahedral chain\tnoteref{t1}}
\tnotetext[t1]{This work was financially supported by the grant of the Slovak Research and Development Agency under the contract No. APVV-16-0186 and by Ministry of Education, Science, Research and Sport of the Slovak Republic under the grant VEGA 1/0043/16.}
 \author{Lucia G\'{a}lisov\'{a}}
 \ead{galisova.lucia@gmail.com}
 \author{Du\v{s}an Kne\v{z}o}
 \address{Institute of Manufacturing Management,
          Faculty of Manufacturing Technologies with a seat in Pre\v{s}ov, \linebreak
          Technical University of Ko\v{s}ice,
          Bayerova 1, 080 01 Pre\v{s}ov, Slovakia}

\begin{abstract}
The ground-state degeneracy and magnetocaloric effect in the spin-$1/2$ Ising-Heisenberg double-tetrahedral chain are exactly investigated. It is demonstrated that the zero-temperature phase diagram involves two classical and two quantum chiral phases with distinct degrees of the macroscopic degeneracy. Different macroscopic degeneracies observed in the latter phases and at individual ground-state phase transitions are confirmed by multiple-peak dependencies of the specific heat and entropy on the magnetic field. The cooling capability of the model is well illustrated by the magnetic-field variations of the isothermal entropy change, temperature isotherms and the magnetic Gr\"uneisen parameter.
\end{abstract}


\end{frontmatter}


\section{Introduction}
\label{sec:1}
The magnetocaloric effect (MCE) has attracted much scientific attention since 1933, when the first successful experiment of the adiabatic demagnetization was performed~\cite{Gia33}. This is mainly due to possible practical applications in magnetic refrigeration technologies and also medicine~\cite{Tis14,Tis16}. The practical use of the MCE is also reflected in an enormous increase in scientific studies devoted to a clarification of the phenomenon both from the experimental and theoretical points of view~\cite{Pech01,Gsc05,Oli10}.

Of particular interest is the investigation of the MCE in one-dimensional (1D) quantum spin models~\cite{Der06,Hon09,Tri10,Top12,Jaf12,Gal14,Str17}. The reason is the possibility to solve they rigorously, as well as, to modify and complement interaction parameters which may be useful for the future quest for novel magnetic materials. Another important aspect is a potential usage of such systems for a correct qualitative explanation of the magnetocaloric measurements realized on real three-dimensional (3D) magnetic compounds~\cite{Sol97,Has08,Kur10,Mat12}. Nowadays, it is known several valuable facts on magnetocaloric properties of the 1D magnetic structures: (i)~significant adiabatic temperature drop can be observed in a vicinity of the field-induced quantum phase transitions and multiple critical points, where more than two ground-state configurations coexist; (ii)~geometric frustration causes a rapid drop of the adiabatic temperature up to the absolute zero; (iii)~rather complex ground-state phase diagram may manifest itself by a sequence of cooling and heating of the system during the adiabatic (de)magnetization; (iv)~even small randomness in the system noticeably diminishes the cooling/heating capability of the system in a proximity of quantum critical points.

\begin{figure}[t!]
\begin{center}
\hspace{0.5cm}
\includegraphics[angle = 0, width = 0.9\columnwidth]{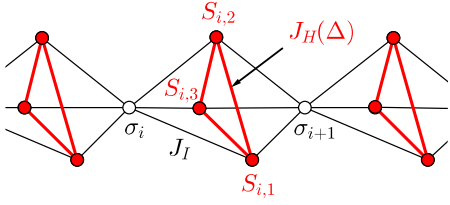}
\vspace{-0.05cm}
\caption{\small A part of the spin-$1/2$ Ising-Heisenberg double-tetrahedral chain.
Thick red lines correspond to the $XXZ$ coupling $J_H(\Delta)$ between the Heisenberg spins at vertices of the triangular clusters (red circles) and thin black lines represent the Ising-type interactions $J_I$ between the Heisenberg spins and the nearest-neighbouring Ising spins at nodal lattice sites (white circles).}
\label{fig1}
\end{center}
\vspace{-0.25cm}
\end{figure}

In this Letter we investigate ground-state degeneracy and magnetocaloric properties of the frustrated spin-$1/2$ double-tetrahedral chain, in which single Ising spins regularly alternate with identical $XXZ$-Heisenberg triangular clusters, as is schematically displayed  in Fig.~\ref{fig1}. The model was originally proposed by V.~Ohanynan {\it et al.}~\cite{Ant09,Oha10} as a simpler variant of the pure spin-$1/2$ Heisenberg chain~\cite{Mam99,Roj03}. It is worthy to note that it belongs to the class of lattice-statistical models with the exact closed-form solution for the partition function at finite temperatures. As has been shown in Refs.~\cite{Ant09,Oha10}, the spin-$1/2$ Ising-Heisenberg double-tetrahedral chain has a rather rich ground-state phase diagram involving several non-chiral and uncommon quantum chiral phases which manifest themselves in an  interesting low-temperature thermodynamics. As a result, the effect of the spin chirality on magnetocaloric properties of the model can be rigorously examined. Another stimulus for examining the aforementioned mixed-spin chain is the copper-based polymeric compound Cu$_3$Mo$_2$O$_9$, which represents the experimental realization of the 1D double-tetrahedral structure~\cite{Has08,Kur10,Mat12}.

The Letter is organized as follows. In Section~\ref{sec:2}, we will briefly describe the magnetic structure of the model and list an exact analytical expression for the Gibbs free energy, which will be subsequently used for exact calculation of basic thermodynamic quantities and some magnetocaloric characteristics. In Section~\ref{sec:3}, we will specify the spin arrangement and macroscopic degeneracy of the ground state. The magnetocaloric properties of the model will also be discussed in Section~\ref{sec:3}. Finally, Section~\ref{sec:4} will summarize the most interesting findings and future outlooks.

\section{Model and its exact treatment}
\label{sec:2}

The spin-$1/2$ Ising-Heisenberg model, which is schematically depicted in Fig.~\ref{fig1}, may also be viewed as a system of $N$ identical bipyramids whose common vertices (white circles) are occupied by the Ising spins and others (red circles) are available for Heisenberg spins. From this perspective, the total Hamiltonian of the model can be written as a sum of $N$ block Hamiltonians, where each block Hamiltonian involves all interaction terms associated with the Heisenberg spin triangle from
one particular bipyramid:
\begin{eqnarray}
\label{eq:H_tot}
{\cal \hat{H}}\!\!\!&=&\!\!\!
\sum_{i = 1}^{N} {\cal \hat{H}}_i,
\\
\label{eq:H_i}
{\cal \hat{H}}_i\!\!\!&=&\!\!\!
J_{H}\sum_{j=1}^{3}
\Big[\Delta\left(\hat{S}_{i,j}^{x}\,\hat{S}_{i,j+1}^{x}\!+\hat{S}_{i,j}^{y}\,\hat{S}_{i,j+1}^{y}\right) + \hat{S}_{i,j}^{z}\,\hat{S}_{i,j+1}^{z}\Big]
\nonumber \\
&+&\!\!\!
J_{I}
\sum_{j=1}^{3}
\hat{S}_{i,j}^{z}\left(\hat{\sigma}_{i}^z + \hat{\sigma}_{i+1}^z\right)
-
\frac{h_I}{2}\left(\hat{\sigma}_{i}^z + \hat{\sigma}_{i+1}^z\right) - h_H\sum_{j=1}^3\hat{S}_{i,j}^{z}.
\nonumber\\
\end{eqnarray}
In above, $\hat{\sigma}_i^z$ and $\hat{S}_{i,j}^{\alpha}$ ($\alpha\in\{x,y,z\}$) represent spatial components of the spin-$1/2$ operators at $i$th nodal lattice site and $j$th vertex of the adjacent $i$th triangular cluster, respectively. To ensure the translational symmetry, the periodic boundary conditions $\hat{\sigma}_{N+1}^z = \hat{\sigma}_{1}^z$ and $\hat{S}_{i,4}^{\alpha} = \hat{S}_{i,1}^{\alpha}$  for the nodal Ising spins and Heisenberg spins, respectively, are imposed. The parameter $J_{H}$ denotes the Heisenberg coupling between spins of the same triangular cluster, and $\Delta$ is the anisotropy parameter, which allows one to control the exchange interaction $J_{H}$ between the easy-axis ($\Delta<1$) and easy-plane ($\Delta>1$) type, as well as, to consider two limiting cases $\Delta=1$ and $\Delta=0$ corresponding to the isotropic Heisenberg and pure Ising models, respectively. The parameter $J_I$ labels the Ising-type exchange interaction between the spins from triangular clusters and their nearest spin neighbours at nodal lattice sites. The last two terms in Eq.~(\ref{eq:H_i}) determine Zeeman energies of the Ising and Heisenberg spins in the longitudinal magnetic field, which may be in general different due to distinct gyromagnetic factors entering into definitions of the 'effective' magnetic fields $h_I$ and $h_H$.

\subsection{Basic thermodynamic quantities}
\label{subsec:2.1}

The quantum mixed spin-$1/2$ double-tetrahedral chain defined by the Hamiltonian~(\ref{eq:H_tot}) can be exactly solved at finite temperatures by two analytical approaches provided the thermodynamic limit $N\to\infty$, namely by the classical transfer-matrix technique~\cite{Kra44,Bax82,Str15} and/or the generalized decoration-iteration mapping transformation~\cite{Fis59,Syo72,Roj09}. Both the methods result in the closed-form analytical expression for the Gibbs free energy ${\cal G}$ of the model:
\begin{equation}
\label{eq:G}
{\cal G} =
N\beta^{-1}\!\ln 2 - N\beta^{-1}\!
\ln\left|W_{1} + W_{-1}
\!+\!\! \sqrt{\left(W_{1} - W_{-1}\right)^2 + 4W_0^2}\,
\right|,
\end{equation}
where $\beta = 1/(k_{\rm B}T)$ ($k_{\rm B}$ is Boltzmann constant, $T$ is the temperature) and $W_{\pm 1}$, $W_{0}$ are functions of the temperature and interaction parameters of the model:
\begin{eqnarray}
W_a \!\!\!&=&\!\!\!  \textrm{e}^{\beta h_Ia/2}
\Bigg\{
\textrm{e}^{-3\beta J_H/4}\cosh\left[\frac{3\beta(J_Ia-h_H)}{2}\right]
\nonumber\\
\!\!\!&&\!\!\!  +
\left[\textrm{e}^{\beta J_H(1-4\Delta)/4} + 2\textrm{e}^{\beta J_H(1+2\Delta)/4}\right]\cosh\left[\frac{\beta(J_Ia  - h_H)}{2}\right]
\Bigg\}.
\nonumber
\end{eqnarray}
The analytical expression~(\ref{eq:G}) allows one to calculate all basic thermodynamic quantities. In particular, the sublattice magnetization $M_I = \langle\hat{\sigma}_i^{z}\rangle$ and $M_{\triangle} = \langle\sum_{j = 1}^3\hat{S}_{i,j}^{z}\rangle$, corresponding to the nodal Ising spin and the Heisenberg triangular cluster, respectively, are uniquely given by the relations:
\begin{equation}
\label{eq:MIMH}
M_I = -\frac{1}{N}\left(\frac{\partial{\cal G}}{\partial h_I}\right)_{T,h_H,\{X\}}, \quad
M_{\triangle} = -\frac{1}{N}\left(\frac{\partial{\cal G}}{\partial h_H}\right)_{T,h_I\{X\}}.
\end{equation}
In above, the set $\{X\}$ includes internal interaction parameters of the model ($\{X\} = \{J_I, J_H,\Delta\}$). In view of above notation, the total magnetization $M$ per elementary unit consisting of one nodal Ising spin and the neighbouring Heisenberg spin triangle can be defined as
\begin{equation}
\label{eq:M}
M = M_I + M_{\triangle}.
\end{equation}
Further, the exact analytical expression for the pair correlation function $C_{II}^{zz} = \langle\hat{\sigma}_i^{z}\sigma_{i+1}^{z}\rangle$, which specify the spin ordering of the Ising spins at $i$th and $(i+1)$st nodal sites, is well known~\cite{Bax82}. Other correlation functions $C_{\triangle}^{zz} = \langle\sum_{j = 1}^3\hat{S}_{i,j}^{z}\hat{S}_{i,j+1}^{z}\rangle$, $C_{\triangle}^{xx} = \langle\sum_{j = 1}^3\hat{S}_{i,j}^{x}\hat{S}_{i,j+1}^{x}\rangle = \langle\sum_{j = 1}^3\hat{S}_{i,j}^{y}\hat{S}_{i,j+1}^{y}\rangle$ and $C_{I\triangle}^{zz} = \langle\hat{\sigma}_i^{z}\sum_{j = 1}^3\hat{S}_{i,j}^{z}\rangle$ follow the relations:
\begin{eqnarray}
\label{eq:Ctriangle}
C_{\triangle}^{zz} \!\!\!&=&\!\!\! \frac{1}{N}\left(\frac{\partial{\cal G}}{\partial J_H}\right)_{J_I,\Delta,\{Y\}}, \quad
C_{\triangle}^{xx} = \frac{1}{N}\left(\frac{\partial{\cal G}}{\partial
\left(J_H\Delta\right)}\right)_{J_I,\{Y\}},
\nonumber\\
C_{I\triangle}^{zz} \!\!\!&=&\!\!\! \frac{1}{2N}\left(\frac{\partial{\cal G}}{\partial
J_I}\right)_{J_H,\Delta,\{Y\}},
\end{eqnarray}
where $\{Y\} = \{T, h_I, h_H\}$. For more computational details, we refer a reader to Ref.~\cite{Ant09}, where the afore-listed physical quantities have been obtained for spin operators with the normalized eigenvalues $\pm1$. Last but not least, for further calculations is valuable to mention even the entropy~${\cal S}$ and the specific heat~${\cal C}$. Both these physical quantities follow from fundamental relations of thermodynamics~\cite{Kit58}:
\begin{equation}
\label{eq:SC}
{\cal S} = -\left(\frac{\partial{\cal G}}{\partial T}\right)_{h_I, h_H, \{X\}}, \quad
{\cal C} = -T\left(\frac{\partial^2{\cal G}}{\partial T^2}\right)_{h_I, h_H, \{X\}}.
\end{equation}

\subsection{Magnetocaloric characteristics}
\label{subsec:2.2}

The knowledge of the thermodynamic quantities listed in Eqs.~(\ref{eq:MIMH})--(\ref{eq:SC}) gives
an opportunity to rigorously determine basic magnetocaloric characteristics, namely the isothermal magnetic entropy change, the isentropic dependence of the temperature
on the magnetic field and also the magnetic Gr\"uneisen parameter. With respect to following discussion, these quantities will be calculated for identical 'effective' magnetic fields acting on the Ising and Heisenberg spins $h_I = h_H = h$.

The isothermal magnetic entropy change $\Delta{\cal S}_{iso}$, which occurs during the isothermal magnetic-field change $\Delta h\!:0\to h^{*}$, can be obtained from the total magnetization data by means of the Maxwell relation~\cite{Pech01}:
\begin{equation}
\label{eq:dSiso1}
\Delta{\cal S}_{iso} = \int_{0}^{h^{*}}\left(\frac{\partial M}{\partial T}\right)_{h,\{X\}}{\rm d}h,
\end{equation}
or as a difference of magnetic entropies at the finite and zero magnetic field under the fixed temperature $T$~\cite{Pech01}:
\begin{equation}
\label{eq:dSiso2}
\Delta{\cal S}_{iso} = \left({\cal S}|_{h = h^{*}} - {\cal S}|_{h = 0}\right)_{T,\{X\}}.
\end{equation}
In the present convention, $-\Delta {\cal S}_{iso}>0$ corresponds to a conventional MCE, while $-\Delta {\cal S}_{iso}<0$ points to an inverse MCE.
The isentropic dependence of the temperature on the magnetic field can be rigorously determined from the transcendent equation
\begin{equation}
\label{eq:Tad}
{\cal S} = \textrm{const.}
\end{equation}
by the bisection method. Finally, the magnetocaloric quantity called the magnetic Gr\"uneisen parameter $\Gamma_h$ can be calculated from the relation~\cite{Zhu03}:
\begin{equation}
\label{eq:Gamma}
{\it\Gamma}_h = -\frac{1}{\cal C}\left(\frac{\partial M}{\partial T}\right)_{h,\{X\}} =  -\frac{1}{T} \frac{\left(\partial {\cal S}/\partial h\right)_{T,\{X\}}}{\left(\partial {\cal S}/\partial T\right)_{h,\{X\}}} = \frac{1}{T}\left(\frac{\partial T}{\partial h}\right)_{{\cal S},\{X\}}.
\end{equation}
Evidently, ${\it\Gamma}_h$ is proportional to the adiabatic cooling rate $\left(\partial T/\partial h\right)_{{\cal S},\{X\}}$, which presents an experimentally measurable quantity~\cite{Lan10,Tok11,Wol11}. Since $\Gamma_h$ is expected to diverge at quantum critical points, it is very useful for an investigation of the cooling efficiency of the model during adiabatic (de)magnetization especially in a vicinity of the field-induced phase transitions.

\section{Results and discussion}
\label{sec:3}

In this section, we discuss a diversity of the ground-state degeneracy, the low-temperature entropy and magnetocaloric properties of the spin-$1/2$ Ising-Heisenberg double-tetrahedral chain defined by the Hamiltonian~(\ref{eq:H_tot}). Taking into account long-standing findings that quantum antiferromagnets exhibit more diverse magnetic behaviour~\cite{Lhu02,Miy03,Die04} and are more efficient low-temperature magnetic coolers than their ferromagnetic counterparts~\cite{Bon72,Bon77}, our analysis is restricted to the case of antiferromagnetic nearest-neighbour couplings $J_I>0$, $J_H>0$ and positive values of the exchange anisotropy parameter $\Delta>0$. Moreover, the number of free parameters is reduced by assuming the identical 'effective' magnetic fields acting on the Ising and Heisenberg spins $h_I = h_H = h$.

\subsection{Ground-state spin arrangement and degeneracy}
\label{subsec:3.1}

\begin{figure}[t!]
\begin{center}
\includegraphics[angle = 0, width = 1.0\columnwidth]{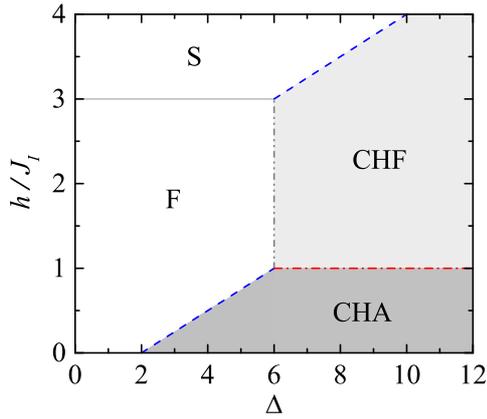}
\vspace{-0.75cm}
\caption{\small The ground-state phase diagram of the spin-1/2 Ising-Heisenberg double-tetrahedral chain in the $\Delta-h$ plane for the interaction ratio $J_H/J_I=0.5$.
The lines of different colors and styles label first-order phase transitions with different macroscopic degeneracies: ${\cal W} = 2^N$ (grey solid line), $3^N$ (blue dashed line), $2^{N+1}$ (grey dot-dot-dashed line), and $\left(3+\sqrt{5}\right)^N$ (red dot-dashed line).}
\label{fig2}
\end{center}
\vspace{-0.25cm}
\end{figure}

First, we examine in detail a ground-state degeneracy of the model. Before doing so, however, it is suitable to briefly recall all possible spin arrangements appearing in the ground state~\cite{Ant09,Oha10}. The zero-temperature parameter space corresponding to the antiferromagnetic version of the model may contain in total four different phases (see Fig.~\ref{fig2}):
\begin{itemize}
\item[(i)]
the saturated (S) phase
\begin{eqnarray}
\label{eq:S}
\hspace{-1.0cm}
      |{\rm S}\rangle = \prod_{i=1}^N
        |\!\uparrow\rangle_{\sigma_i}\!\otimes\!
        |\uparrow\uparrow\uparrow\rangle_{\triangle_i},
\end{eqnarray}
with the energy $E^{\rm S}= \dfrac{N}{4}\left(3J_H + 6J_I - 8h\right),$
\item[(ii)]
the ferrimagnetic (F) phase
\begin{eqnarray}
\label{eq:F}
\hspace{-1.0cm}
|{\rm F}\rangle = \prod_{i=1}^N
        \begin{cases}
        \lefteqn{
        |\!\uparrow\rangle_{\sigma_i}\!\otimes\!
        |\downarrow\downarrow\downarrow\rangle_{\triangle_i} \qquad  (h=0)
                }
                \\[1mm]
       \lefteqn{
        |\!\downarrow\rangle_{\sigma_i}\!\otimes\!
        |\uparrow\uparrow\uparrow\rangle_{\triangle_i} \qquad  (h\geq0)
                }
        \end{cases}
\end{eqnarray}
with the energy $E^{\rm F}= \dfrac{N}{4}\left(3J_H - 6J_I - 4h\right),$
\item[(iii)]
the chiral ferrimagnetic (CHF) phase
\begin{eqnarray}
\label{eq:Fch}
\hspace{-1.0cm}
      |{\rm CHF}\rangle = \prod_{i=1}^N
      |\!\uparrow\rangle_{\sigma_i}\!\otimes\!
       \frac{1}{\sqrt{3}}\left(
        |\uparrow\uparrow\downarrow\rangle
       +\omega^{\pm}|\uparrow\downarrow\uparrow\rangle
       +\omega^{\mp}|\downarrow\uparrow\uparrow\rangle
        \right)_{\triangle_i}
\end{eqnarray}
with the energy $E^{\rm CHF}= -\dfrac{N}{4}\left(J_H+2J_H\Delta - 2J_I + 4h\right),$
\item[(iv)]
the chiral antiferromagnetic (CHA) phase
\begin{eqnarray}
\label{eq:AFch}
\hspace{-1.0cm}
      |{\rm CHA}\rangle = \prod_{i=1}^N
        \begin{cases}
        \lefteqn{
        |\!\uparrow\rangle_{\sigma_i}\!\otimes\!
        \frac{1}{\sqrt{3}}\left(
        |\downarrow\downarrow\uparrow\rangle
       +\omega^{\pm}|\downarrow\uparrow\downarrow\rangle
       +\omega^{\mp}|\uparrow\downarrow\downarrow\rangle
        \right)_{\triangle_i}
                }
                \\[3mm]
       \lefteqn{
        |\!\downarrow\rangle_{\sigma_i}\!\otimes\!
        \frac{1}{\sqrt{3}}\left(
        |\uparrow\uparrow\downarrow\rangle
       +\omega^{\pm}|\uparrow\downarrow\uparrow\rangle
       +\omega^{\mp}|\downarrow\uparrow\uparrow\rangle
        \right)_{\triangle_i}
                }
        \end{cases}
\end{eqnarray}
with the energy $E^{\rm CHA}= -\dfrac{N}{4}\left(J_H+2J_H\Delta + 2J_I\right).$
\end{itemize}
In above, the product $\prod_{i=1}^N$ runs over elementary units forming double-tetrahedrons and $\omega^{\pm} = {\rm e}^{\pm 2\pi{\rm i}/3}$ (${\rm i}^2 = -1$). The single-site ket vectors determine the up or down state of the Ising spin at $i$th nodal lattice site, while other three-site ones refer to the spin arrangement peculiar to the neighbouring (also $i$th) Heisenberg spin triangle.

Obviously, the S phase has a unique spin ordering due to a full polarization of all spins into the magnetic-field direction. On the other hand, if the classical F phase contains the ground state, the Ising spins may occupy either the up or down state with the same probability at $h=0$. The Heisenberg spins in triangular clusters also choose between two possible classical ferromagnetic arrangements peculiar to the total spins $S_{tot}^z = -3/2$ and $3/2$ in order to preserve the spontaneous antiferromagnetic order with their Ising neighbours. As expected, the two-fold degeneracy of the F phase is canceled as soon as the external magnetic field is switched on.
The latter two quantum CHF and CHA phases are frustrated (macroscopically degenerated) in whole their stability regions due to two possible chiral degrees of freedom of each Heisenberg triangle. Moreover, the CHA phase is degenerated even in terms of two energetically equivalent spin configurations, which are characterized by opposite signs of $\sigma_i^z$ and $S_{tot}^z$ corresponding to the nodal Ising spins and Heisenberg triangular clusters, respectively. Thus, the macroscopic degeneracy of the frustrated CHA phase is two-times higher than that one of the frustrated CHF phase, which equals to ${\cal W} = 2^N$.
The spin arrangement inherent to the individual ground-state phases~(\ref{eq:S})--(\ref{eq:AFch}) can be independently confirmed by the asymptotic values of the sublattice magnetization and correlation functions listed in Table~\ref{tab:1}, which have been calculated from Eqs.~(\ref{eq:MIMH}) and (\ref{eq:Ctriangle}) in the zero-temperature limit $T\to 0$.

Other degeneracies in the zero-temperature $\Delta-h$ parameter plane are apparent from the caption of Fig.~\ref{fig2}. Obviously, the field-induced phase boundary between the classical F and S phases has the macroscopic degeneracy of the same size as the quantum CHF phase. Its origin is, however, quite different. The degeneracy ${\cal W} = 2^N$ observed at the saturation field $h/J_I = 3$, where the system passes from the phase F to the S one, comes merely from zero-temperature spin fluctuations at nodal lattice sites between up and down states which can be detected in neighbouring phases [see Eqs.~(\ref{eq:S}) and~(\ref{eq:F})].
\begin{table}[t!]
\vspace{-0.0cm}
\caption{\small The zero-temperature values of the sublattice magnetization and pair correlation functions inherent to relevant ground-state phases.}
\label{tab:1}
\centering
\begin{tabular*}{1.0\columnwidth}{@{\vrule height 5pt depth 2pt width0pt\extracolsep\fill
}lccccccl}
\hline
  GS phases & $M_I$ & $M_{\triangle}$ & $C_{II}^{zz}$ & $C_{\triangle}^{zz}$ & $C_{\triangle}^{xx}$ & $C_{I\triangle}^{zz}$\\[0.3mm]
\hline\hline
  S & $1/2$ & $3/2$ & $1/4$ & $3/4$ & $0$ & $3/4$ \\
  F \,\,(at $h=0$)& $0$ & $0$ & $1/4$ & $3/4$ & $0$ & $-3/4$ \\
  F \,\,(at $h>0$)& $-1/2$ & $3/2$ & $1/4$ & $3/4$ & $0$ & $-3/4$ \\
  CHF & $1/2$ & $1/2$ & $1/4$ & $-1/4$ & $-1/4$ & $1/4$ \\
 CHA & $0$ & $0$ & $1/4$ & $-1/4$ & $-1/4$ & $-1/4$\\
\hline
\end{tabular*}
\end{table}
The degeneracies identified at other ground-state boundaries, that separate chiral phase from non-chiral one and two different chiral phases, are significantly higher, namely ${\cal W} = 3^N$, $2^{N+1}$ and $\left(3+\!\!\sqrt{5}\right)^{N}$. The first degeneracy can be found at the field-induced phase transitions CHA--F, CHF--S and is associated with a mutual coexistence of classical fully saturated states and quantum chiral states of the Heisenberg three-site clusters. The second one appears at the phase transition along field-independent line separating the F phase from the CHF one due to zero-temperature spin fluctuations of both the Ising and Heisenberg spins. Finally, the last (highest) macroscopic degeneracy ${\cal W} = \left(3+\!\!\sqrt{5}\right)^{N}$ can be observed at the field-induced phase boundary CHA--CHF. It can be easily proved that this highly non-trivial value is a result of the mutual coexistence of the neighbouring phases CHF, CHA with another (novel) chiral ferrimagnetic (CHF$^{*}_{i}$) double-tetrahedral spin configuration with the antiferromagnetic alignment of the Ising spins at $i$th and $(i+1)$st nodal lattice sites, which is given by the following eigenvector and ground-state energy:
\begin{eqnarray}
\hspace{-0.75cm}
\label{eq:NAFch}
|{\rm CHF^{*}}\rangle_i \!\!\!\!&=&\!\!\! |\!\uparrow\rangle_{\sigma_i}\!\otimes\!
       \frac{1}{\sqrt{3}}\left(
        |\uparrow\uparrow\downarrow\rangle
       +\omega^{\pm}|\uparrow\downarrow\uparrow\rangle
       +\omega^{\mp}|\downarrow\uparrow\uparrow\rangle
        \right)_{\triangle_i}\!\otimes\!|\!\downarrow\rangle_{\sigma_{i+1}}
\nonumber\\
\hspace{-0.75cm}
E^{{\rm CHF^{*}}}_{i} \!\!\!\!&=&\!\!\! -\dfrac{J_H}{4} -\dfrac{J_H\Delta}{2} +\dfrac{J_I}{2} -\dfrac{h}{2}.
\end{eqnarray}
For computational details, we refer the reader to appendices of the recent works~\cite{Tor16, Tor18}, where the calculation of the macroscopic degeneracy of the triple point, at which three ground states coexist together with another spin configuration with the equal energy, is presented in detail.

Before concluding the subsection, it is noteworthy that the degeneracies of the classical F phase and the frustrated CHA phase were not discussed in the original works~\cite{Ant09,Oha10}, because their omission did not affect presented results. On the other hand, the both ones, as well as the field-independent chirality of triangular clusters present in CHF and CHA phases, have been analyzed in detail in our recent works~\cite{Gal17,Gal18}, which deal with the hybrid spin-electron double-tetrahedral chain. Moreover, the macroscopic degeneracies ${\cal W} = 2^N$ and $3^N$ observed at the phase transitions F--S and CHA--F, CHF--S, respectively, have also been confirmed at the field-induced ground-state phase transitions of the same kind in the latter work~\cite{Gal18}. Thus, it seems that the degeneracies identified at individual ground states and field-induced phase transitions are a general feature of the 1D lattice-statistical models with the elementary unit cell of trigonal bipyramids.

\subsection{Entropy and specific heat}
\label{subsec:3.2}

\begin{figure}[t!]
\begin{center}
\includegraphics[angle = 0, width = 1.0\columnwidth]{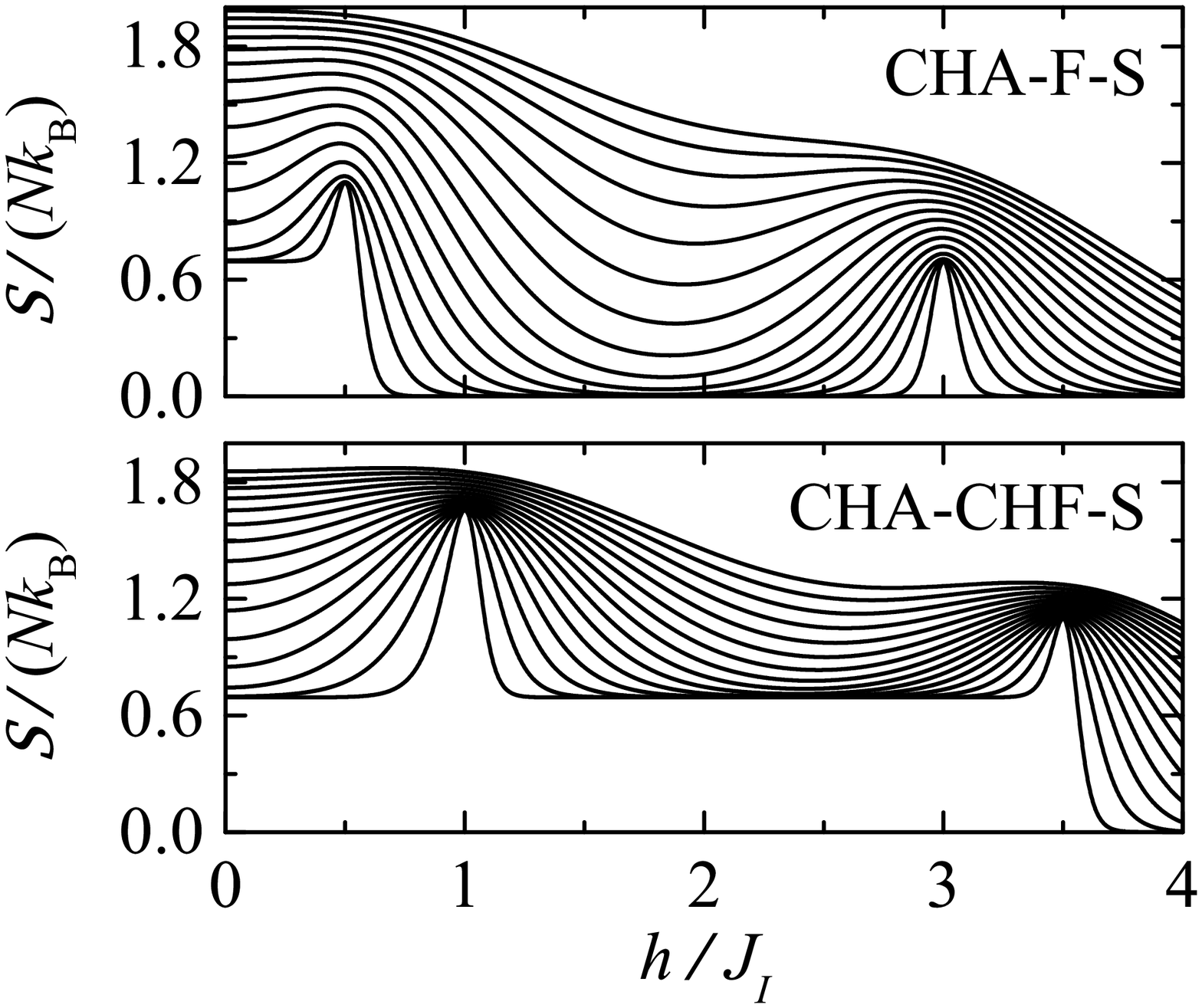}
\vspace{-0.75cm}
\caption{\small The entropy per elementary unit as a function of the magnetic field for the interaction ratio $J_H/J_I=0.5$ and two representative values of the exchange anisotropy $\Delta=4$ (upper panel), $\Delta=8$ (lower panel) at the temperatures $k_{\rm B}T/J_I = 0.03,0.06,\ldots,0.45$ (from bottom to top in both panels).}
\label{fig3}
\end{center}
\vspace{0.25cm}
\begin{center}
\includegraphics[angle = 0, width = 1.0\columnwidth]{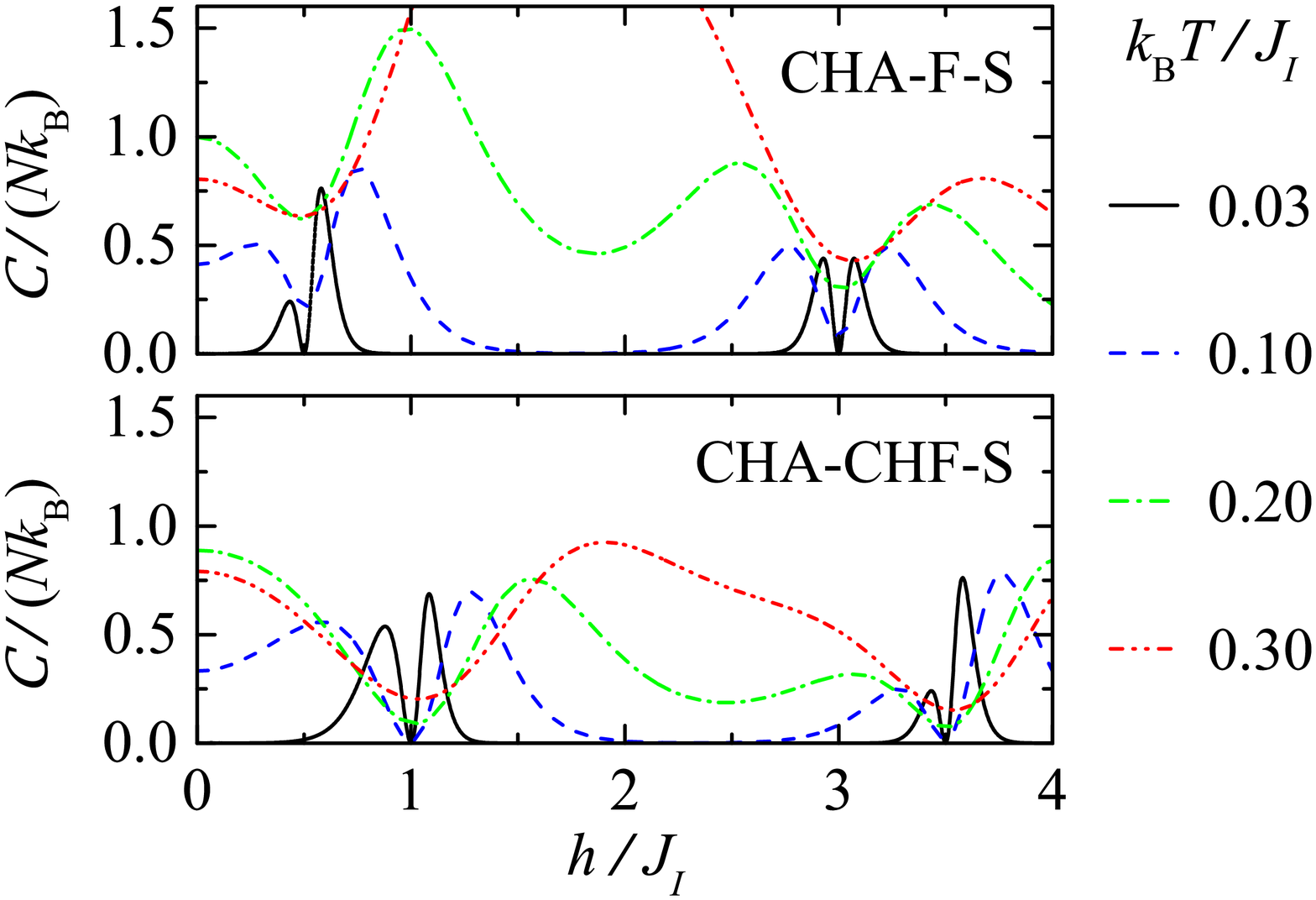}
\vspace{-0.75cm}
\caption{\small The specific heat per elementary unit as a function of the magnetic field for the same values of the interaction parameters as in Fig.~\ref{fig3} by considering different temperatures.}
\label{fig4}
\end{center}
\vspace{-0.5cm}
\end{figure}

The macroscopic ground-state degeneracy of the frustrated mixed-spin chain under investigation manifests itself in abrupt low-temperature variations of the entropy and specific heat. The entropy value per one four-spin unit, which consists of one nodal Ising spin and three Heisenberg spins from the neighbouring triangular cluster, can be directly determined from the degeneracy of the ground-state manifold according to the formula ${\cal S}/(Nk_{\rm B}) = \lim\limits_{N\to\infty}N^{-1}\ln{\cal W}$~\cite{Der04}.
It follows from previous discussion that the ground-state entropy normalized per unit cell is zero only at the classical F and S phases. Otherwise, ${\cal S}/(Nk_{\rm B})$ takes finite values. In particular, if the quantum phases CHF, CHA constitute the ground state, the zero-temperature entropy reaches the value ${\cal S}/(Nk_{\rm B}) = \ln 2 \approx0.693$ due to two possible chiral degrees of freedom of each Heisenberg spin triangle. Moreover, the residual entropy can also be observed at individual ground-state phase transitions between neighbouring phases, namely ${\cal S}/(Nk_{\rm B}) = \ln 2\approx0.693$ at F--S, F--CHF, ${\cal S}/(Nk_{\rm B}) = \ln 3\approx1.099$ at CHA--F, CHF--S, and ${\cal S}/(Nk_{\rm B}) = \ln\left(3+\sqrt{5}\right)\approx1.656$ at CHA--CHF.
The aforementioned statements may be independently confirmed by isothermal entropy dependencies displayed in Fig.~\ref{fig3}, which have been numerically obtained from Eq.~(\ref{eq:SC}). Clearly, the ${\cal S}(h)$ curves plotted for low enough temperature $k_{\rm B}T/J_I = 0.03$ exhibit narrow peaks at the critical fields which correspond to the first-order phase transitions. As expected, magnitudes of the peaks coincide with previously listed values of the normalized residual entropy. The zero entropy between different peaks is pertinent to the uniquely ordered phases F and S, whereas the plateau at ${\cal S}/(Nk_{\rm B}) = \ln 2$ corresponds to the macroscopically degenerate chiral phases CHA and CHF.

Fig.~\ref{fig4} displays magnetic-field variations of the specific heat for a few different temperatures and the same values of the interaction parameters $J_H/J_I$, $\Delta$ as in Fig.~\ref{fig3}.
As can be seen, the low-temperature ${\cal C}(h)$ curves have a multi-peak structure with two local maxima symmetrically located around each critical field.
The occurring specific-heat maxima are the result of strong thermal excitations between ground-state spin configuration and low-lying excited state with the spin arrangement peculiar to the neighbouring ground state.
It should be noted that the heights of most peaks can be well explained by the Schottky theory for a two-level system~\cite{Gop66}. Specifically, the specific-heat peaks which arise in a vicinity of the critical field corresponding to the ground-state phase transition F--S are of the height ${\cal C}^{max}/(Nk_{\rm B}) \approx 0.439$, because both the ground state and the first excited state are uniquely ordered (non-degenerated) around this particular phase boundary (see the peaks in a vicinity of the critical field $h_c/J = 3$ in upper panel of Fig~\ref{fig4}).
On the other hand, the Schottky-type peaks of two different heights ${\cal C}^{max}/(Nk_{\rm B}) \approx 0.241$ and $0.762$ can be identified nearby the field-induced phase transitions CHA--F and CHF--S (see the peaks around the critical field $h_c/J = 0.5$ in upper panel and $h_c/J = 3.5$ in lower panel of Fig~\ref{fig4}). The lower one is appeared slightly below the appropriate phase transition, where the ground state is frustrated due to chiral degrees of freedom of the Heisenberg spin triangles, while the low-lying excited state is macroscopically non-degenerate.
The higher one can be observed above the phase transition owing to the reversed degeneracy scenario at zero temperature  and in the first excited state.
Finally, it is worthy to mention two Schottky-type maxima ${\cal C}^{max}/(Nk_{\rm B}) \simeq 0.538$ and $0.688$, which arise in the low-temperature specific heat curves ${\cal C}(h)$ nearby the critical field $h_c/J = 1$ at sufficiently strong exchange anisotropy $\Delta>4\left(J_H/J_I\right)^{-1} - 2$ (see lower panel of Fig~\ref{fig4}). We highlight that these peaks are located nearby the ground-state phase transition at which two neighbouring phases CHA, CHF are in thermodynamic equilibrium with third chiral double-tetrahedral spin configuration CHF$^{*}$ defined by Eq.~(\ref{eq:NAFch}). Therefore, one has to employ more general three-level model approach~\cite{Mei73} to describe their actual heights instead of two-level one.

\subsection{Enghanced MCE}
\label{subsec:3.3}

In this part, we turn our discussion to the magnetocaloric properties of the studied spin-$1/2$ Ising-Heisenberg double-tetrahedral chain.

\begin{figure}[b!]
\begin{center}
\includegraphics[angle = 0, width = 1.0\columnwidth]{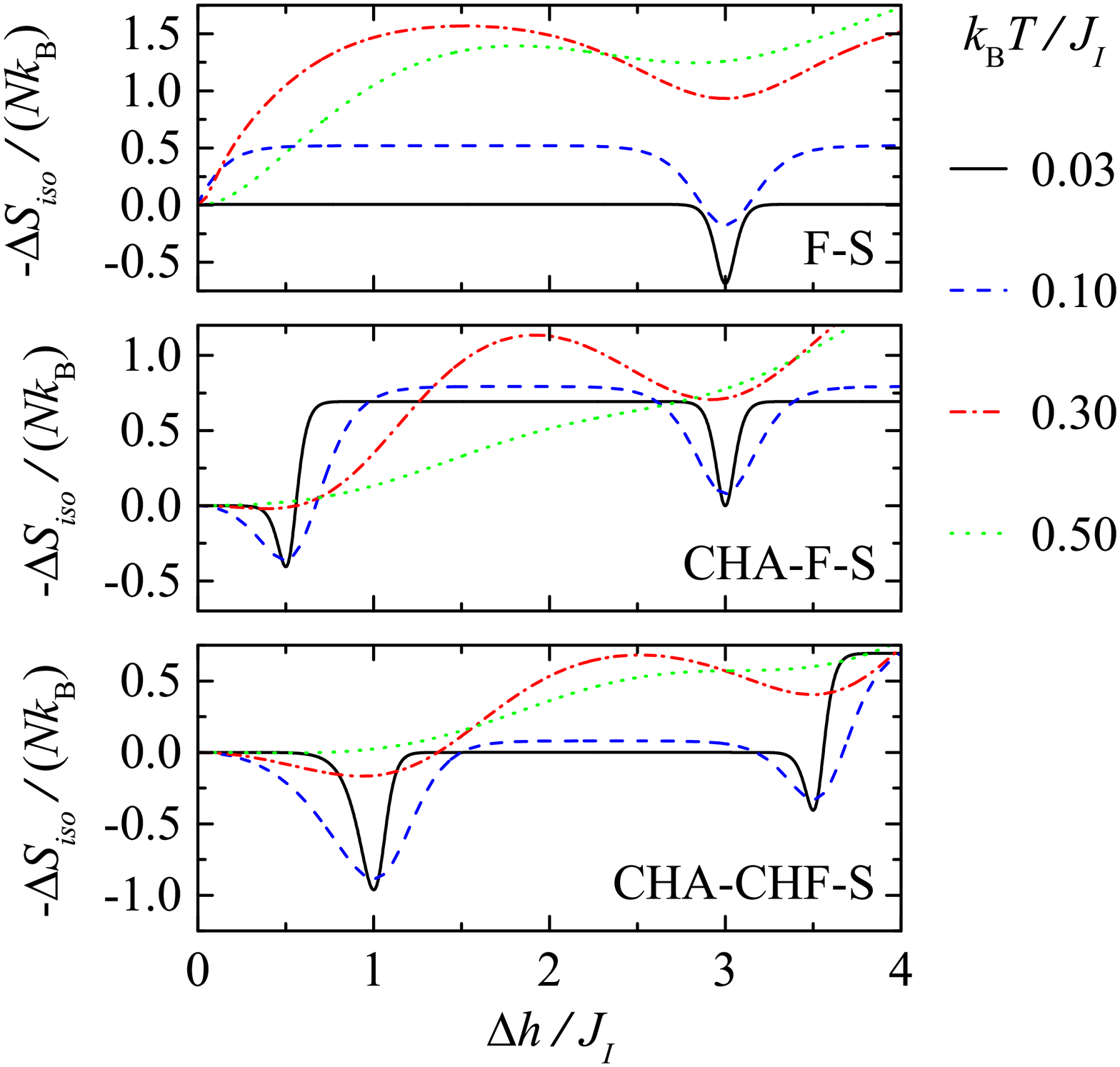}
\vspace{-0.75cm}
\caption{\small The isothermal entropy change per elementary unit as a function of the magnetic-field change $\Delta h\!:0\to h$ for the interaction ratio $J_H/J_I=0.5$ and three representative values of the exchange anisotropy $\Delta=1$ (upper panel), $\Delta=4$ (middle panel), $\Delta=8$ (lower panel) at different temperatures.}
\label{fig5}
\end{center}
\vspace{-0.25cm}
\end{figure}

First we start with examining the MCE during the isothermal (de)magnetization process. For this purpose, several magnetic-field dependencies of the isothermal entropy change are displayed in Fig.~\ref{fig5} for three representative values of the exchange anisotropy at a few different temperatures.
Evidently, the low-temperature thermal change of the entropy is zero in those magnetic-field regions, where the ground-state degeneracies at finite and zero magnetic field are equal. On the contrary, if the finite-field degeneracy of the ground state is different from the zero-field one, $-\Delta{\cal S}_{iso}$ reaches a finite value. In particular, $-\Delta{\cal S}_{iso}/(Nk_{\rm B}) \approx 0.693$ can be found in the magnetic-field ranges pertinent to the ordered F and S phases in the temperature limit $T\to0$ when the frustrated CHA phase constitutes the zero-field ground state. This plateau clearly points to the conventional MCE in both the phases.
It is obvious from Fig.~\ref{fig5} that the plateaus observed at $-\Delta{\cal S}_{iso}/(Nk_{\rm B}) \approx 0$ and $0.693$ are separated by pronounced minima, which arise at the critical fields corresponding to the ground-state phase transitions. Their existence clearly point out to a rapid change in magnetocaloric properties of the model at these particular points. The depths of individual minima are uniquely given by a difference between degeneracies (entropies) of the zero-field ground state and  appropriate first-order phase transition.
As expected, the increasing temperature causes the gradual lifting and smoothing of the minima as well as the destruction of the plateaus due to the increasing thermal spin fluctuations in the system.

\begin{figure}[t!]
\begin{center}
\includegraphics[angle = 0, width = 1.0\columnwidth]{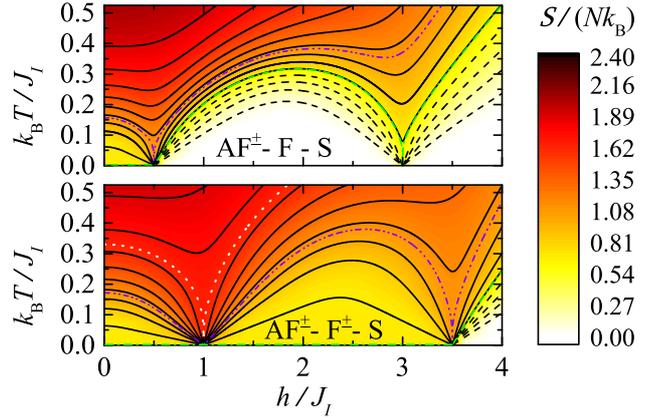}
\vspace{-0.75cm}
\caption{\small The entropy density plots in the $h-T$ parameter plane for the same values of $J_H/J_I$ and $\Delta$ as in Fig.~\ref{fig3}. The curves correspond to the isentropic temperature changes at the residual entropies ${\cal S}/(Nk_{\rm B}) = 0.01, 0.25, 0.4, 0.55$ (dashed lines), ${\cal S}/(Nk_{\rm B}) = 0.7, 0.85, 1, \ldots, 1.9$ (solid lines), ${\cal S}/(Nk_{\rm B}) = \ln 2$ (dash-dotted line), ${\cal S}/(Nk_{\rm B}) = \ln 3$ (dot-dot-dashed line), and ${\cal S}/(Nk_{\rm B}) = \ln\left(3+\!\!\sqrt{5}\right)$ (dotted line).}
\label{fig6}
\end{center}
\vspace{-0.25cm}
\end{figure}

The temperature response of the model to the magnetic-field changes under the adiabatic condition~(\ref{eq:Tad}) can be well understood from Fig.~\ref{fig6}, which shows the entropy density plots in the $h-T$ parameter plane along with temperature dependencies of the model on the magnetic field at various entropies. For easy reference, the interaction parameters of the model are chosen as in Fig.~\ref{fig3}.
As evidenced from Fig.~\ref{fig6}, the isentropic dependencies of the temperature on the magnetic field show significant changes near the critical fields whenever the macroscopic degeneracy of the system is close enough to the ground-state degeneracies at these particular points. The rapid decrease (increase) in temperature indicates an enhancement of the MCE during the adiabatic (de)magnetization process in these regions. It should be noted, however, that the system cools down up to the zero temperature solely when entropy values are lower than or equal to that ones at the appropriate critical fields.
In accordance to the previous discussion on the ground-state entropy (subsection~\ref{subsec:3.2}), the fastest adiabatic cooling of the model up to the absolute zero temperature can be observed for three residual entropies: ${\cal S}/(Nk_{\rm B}) = \ln 2$, when $h$ approaches the value corresponding to the F--S phase boundary, ${\cal S}/(Nk_{\rm B}) = \ln 3$, when $h$ tends to the values pertinent to the ground-state phase transitions CHA--F and CHF--S, and ${\cal S}/(Nk_{\rm B}) = \ln \left(3+\!\!\sqrt{5}\right)$, when $h$ approaches the value corresponding to the  CHA--CHF phase transition.

\begin{figure}[t!]
\begin{center}
\includegraphics[angle = 0, width = 1.0\columnwidth]{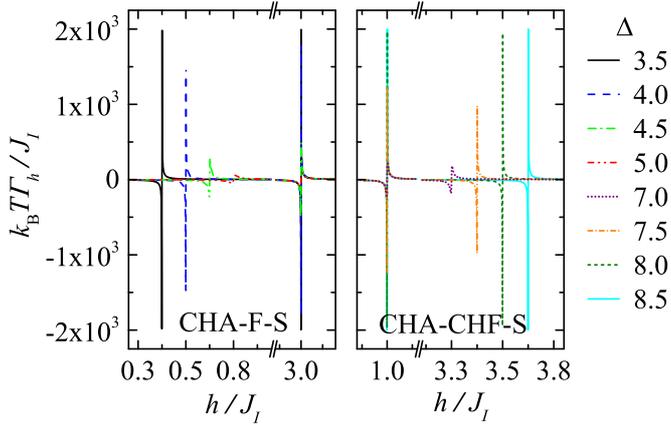}
\vspace{-0.75cm}
\caption{\small The cooling rate versus magnetic field for the interaction ratio $J_H/J_I = 0.5$ and a few different values of the exchange anisotropy $\Delta$ at the fixed temperature $k_{\rm B}T/J_I = 0.03$.}
\label{fig7}
\end{center}
\vspace{-0.25cm}
\end{figure}

To clarify the cooling rate of the model nearby individual field-induced ground-state phase transitions in dependence on internal parameters of the model, the magnetic Gr\"uneisen parameter multiplied by the temperature versus magnetic field is plotted in Fig.~\ref{fig7} for the interaction ratio $J_H/J_I=0.5$ and a few different values of the exchange anisotropy $\Delta$ at the fixed temperature $k_{\rm B}T/J_I = 0.03$ (the lowest possible temperature for numerical calculations).
As one can see, the depicted dependencies of $k_{\rm B}T{\it\Gamma}_h/J_I$ exhibit very pronounced sharp local maxima (minima) slightly above (below) respective ground-state phase transitions.
The observed peaks gradually decrease as the exchange anisotropy approaches the value which is associated with the zero-temperature transition between the classical and quantum chiral ferrimagnetic phases (compare Fig.~\ref{fig7} with the ground-state phase diagram shown in Fig.~\ref{fig2}).
One can thus conclude that the adiabatic cooling capability of the system observed nearby the field-induced phase transitions generally decreases with approaching to the  F--CHF phase boundary.
It is worthy to note that the same scenario can be observed for the fixed exchange anisotropy $\Delta$ and varying interaction ratio $J_H/J_I$. Considering this fact, the figure showing the low-temperature dependencies of $k_{\rm B}T{\it\Gamma}_h/J_I$ versus magnetic field for the fixed $\Delta$ and different values of $J_H/J_I$ along with the associated discussion are omitted from this Letter without loss of completeness.

\section{Summary and future outlooks}
\label{sec:4}

The present Letter deals with the ground-state and magnetocaloric properties of the frustrated spin-$1/2$ double-tetrahedral chain whose nodal lattice sites occupied by the Ising spins regularly alternate with identical $XXZ$-Heisenberg triangular clusters.
Using the exact solution presented in Refs.~\cite{Ant09,Oha10}, we have exactly calculated the basic thermodynamic quantities, such as the Gibbs free energy, the total and sublattice magnetization, pair correlation functions, the entropy, and the specific heat. Their detailed examination has shown that the ground-state diagram involves the common saturated phase, the classical ferrimagnetic phase showing two-fold degeneracy at zero magnetic field, and two quantum frustrated phases CHF, CHA whose field-independent macroscopic degeneracy is caused by two possible chiral degrees of freedom of each Heisenberg triangular cluster.

We have demonstrated that the frustration observed in the quantum CHA phase and various macroscopic degeneracies at individual field-induced ground-state phase transitions are manifested themselves in the enhanced MCE of different intensities during the isothermal and adiabatic (de)magnetization process. Namely, significant local minima arising in the low-temperature $-\Delta{\cal S}_{iso}(h)$ curves at critical fields, which correspond to relevant ground-state phase transitions, clearly point out to a rapid change in magnetocaloric properties of the model at these particular points. Moreover, the finite plateau $-\Delta{\cal S}_{iso}/(Nk_{\rm B}) \approx 0.693$ of the isothermal entropy change, which can be observed in the magnetic-field ranges corresponding to the F and/or S phases, clearly points to the conventional MCE in both the phases when the zero-field ground state is frustrated due to the spin chirality of the Heisenberg three-site clusters. Furthermore, it has been evidenced that the adiabatic cooling capability of the studied 1D mixed-spin model is the most pronounced when the macroscopic degeneracy is close enough to the ground-state degeneracies observed at individual ground-state phase transitions. However, the adiabatic cooling rate rapidly decreases with approaching to the F--CHF boundary.

Finally, it should be mentioned that the investigated spin-$1/2$ Ising-Heisenberg double-tetrahedral chain can be very easily extended to a higher dimension, since it belongs to a prominent class of bond-decorated lattice-statistical models~\cite{Bax82, Str15}. Our future work will continue in this direction.


\end{document}